\documentclass{article}

\pdfoutput=1

\usepackage{fullpage}
\RequirePackage[OT1]{fontenc}
\RequirePackage{amsthm,amsmath,amssymb,mathtools}
\RequirePackage{natbib}
\RequirePackage[colorlinks,citecolor=blue,urlcolor=blue]{hyperref}
\RequirePackage[usenames,dvipsnames]{xcolor}
\RequirePackage{graphicx,subcaption}
\RequirePackage{ulem}

\newcommand{\E}{\mathbb{E}}

\begin{document}

\title{Sensitivity analysis for an unobserved moderator in RCT-to-target-population generalization of treatment effects}

\author{Trang Quynh Nguyen, Cyrus Ebnesajjad, Stephen R. Cole, Elizabeth A. Stuart}

\maketitle

\begin{abstract}
In the presence of treatment effect heterogeneity, the average treatment effect (ATE) in a randomized controlled trial (RCT) may differ from the average effect of the same treatment if applied to a target population of interest.
If all treatment effect moderators are observed in the RCT and in a dataset representing the target population, we can obtain an estimate for  the target population ATE by adjusting for the difference in the distribution of the moderators between the two samples.
This paper considers sensitivity analyses for two situations: (1) where we cannot adjust for a specific moderator $V$ observed in the RCT because we do not observe it in the target population; and (2) where we are concerned that the treatment effect may be moderated by factors not observed even in the RCT, which we represent as a composite moderator $U$.
In both situations, the outcome is not observed in the target population.
For situation (1), we offer three sensitivity analysis methods based on (i) an outcome model, (ii) full weighting adjustment, and (iii) partial weighting combined with an outcome model.
For situation (2), we offer two sensitivity analyses based on (iv) a bias formula and (v) partial weighting combined with a bias formula.
We apply methods (i) and (iii) to an example where the interest is to generalize from a smoking cessation RCT conducted with participants of alcohol/illicit drug use treatment programs to the target population of people who seek treatment for alcohol/illicit drug use in the US who are also cigarette smokers.
In this case a treatment effect moderator is observed in the RCT but not in the target population dataset.


\end{abstract}

\section{Introduction}

Randomized controlled trials (RCTs) can be used to obtain unbiased estimates of the effect of the intervention of interest in the sample used in the trial, resulting in high \textit{internal validity}.
However, standard RCTs are not necessarily informative regarding the effects an intervention would have in a target population that may be somewhat different from the RCT sample; in other words, the RCT may have limited \textit{external validity} or \textit{generalizability}.
Potential challenges in drawing inferences for populations of policy or decision-making relevance are becoming an increasing concern, as researchers aim to make their research results as relevant as possible. 

As shown by \citet{Weisberg2009}, \citet{Cole2010} and \citet{Olsen2013a}, results from RCTs may not directly carry over to populations if there are treatment effect moderators whose distribution differs between the RCT sample and the target population.
Methods for assessing \citep{Greenhouse2008,Stuart2011,Stuart2015} and enhancing \citep{Cole2010,Tipton2013,Kern2016a} generalizability have been proposed. The latter includes approaches that reweight the RCT sample so that it resembles the target population with respect to the observed covariates and plausible moderators \citep{Cole2010,Kern2016a} or predict treatment effects for target population members based on an outcome model that captures effect heterogeneity \citep{Kern2016a}.
However, those methods only adjust for observed characteristics.
In practice, once a dataset is identified as representing the target population, it is often found that the number of variables measured consistently between this dataset and the RCT is small \citep*[][in press]{Stuart2015,Stuart}.
In many cases, researchers and policymakers may be worried about unobserved differences between the RCT sample and the target population and how much they influence the conclusions regarding population effects. 

This paper presents a set of approaches for assessing the sensitivity of population effect estimates to unobserved moderators, to be used when generalizing from a RCT to a target population.
These sensitivity analyses are analogous to methods that assess sensitivity to an unobserved confounder in observational studies \citep*[such as][]{Cornfield1959,Rosenbaum1983,Rosenbaum1987,Gastwirth1998,Greenland1996,Schneeweiss2006,Arah2008,Vanderweele2011a,Ding2014,Ding2016}.
They address two situations: (1) when a specific treatment effect moderator is observed in the RCT but is not measured in the target population; and (2) when researchers are concerned about possible effect moderation by factors that are not observed even in the RCT.

The data application in this paper involves generalizing the effect of a smoking cessation intervention from a RCT conducted with participants in alcohol/illicit drug use treatment programs \citep{Reid2008} to the target population of people who seek treatment for alcohol/illicit drug use in the US who are also cigarette smokers.
This RCT is one of the substance use treatment RCTs funded by the US National Institute on Drug Abuse (NIDA); these are deposited in a repository maintained by NIDA's Clinical Trials Network with the purpose to facilitate the use of evidence from RCTs to generate knowledge that informs the provision of treatment services to people with substance use disorders in the US.

With a subset of these RCTs (not including the one in our current application), Susukida and colleagues found significant differences in certain characteristics between the RCT samples and samples they identified as representing relevant target populations \citep{Susukida2016}, and for some interventions, a substantial difference between the average treatment effects (ATEs) for the target population and for the RCT sample due to treatment effect heterogeneity associated with such characteristics (work under review).
Such work considers only variables measured in both each RCT and the corresponding target population dataset.
With the proposed sensitivity analysis methods, we are able to take one step further, exploring treatment effect moderators among all baseline variables measured in the RCT and conducting sensitivity analysis when finding that one moderator (baseline cigarette addiction score) is not observed in the target population dataset (here drawn from the National Survey on Drug Use and Health, or NSDUH).

The paper is structured as follows: Section \ref{sxn:Z} describes two methods for obtaining estimates for target population treatment effects when the moderators are observed in both the RCT and the target population dataset; these are the basis of the sensitivity analyses we propose.
Section \ref{sxn:V} presents sensitivity analysis methods for settings where a moderator is observed in the RCT but not in the target population.
Section \ref{sxn:U} addresses sensitivity analyses for effect moderation that is not even observed in the RCT.
Section \ref{sxn:application} reports on the data application.
Section \ref{sxn:discuss} concludes with a discussion.

\section{Two methods for generalization when the moderators are observed in both the RCT and in a dataset representing the target population}\label{sxn:Z}

This section formalizes the goal of inference, desbribes notation, and reviews two methods for generalizing treatment effect estimates from an RCT to a target population; these methods form the basis for the sensitivity analyses described below.

Consider a RCT in which participants are randomly assigned to active treatment ($T=1$) and control ($T=0$) conditions, and their outcomes ($Y$) are observed.
In this sample, we also observe pre-treatment covariates, including covariates $Z$ that interact with treatment in influencing the outcome, and covariates $X$ that influence the outcome but do not interact with treatment. $Z$ and $X$ are generally multivariate, but we use univariate notation to simplify presentation.

Suppose we have data from a second sample, one that represents the target population. Let $S$ represent sample membership, with $S=1$ if a person is in the RCT and $S=0$ if a person is in the target population sample.
Here we assume that the two samples are disjoint. (For the case where the RCT sample is a subset of the target population sample, the methods are slightly modified \citep*{Cole2010}, which we comment on in the Discussion section.)
In this section, we consider the situation where we also observe the treatment effect moderators $Z$ in the target population dataset.
All through this paper we assume that the outcome is not observed in the target population.

Let $Y^t$ denote the potential outcome if under treatment condition $t,t\in\{0,1\}$.
For each RCT participant, we observe one of the two potential outcomes $Y^1,Y^0$.
For those in the target population sample, we observe neither.
We are interested in the average treatment effects (ATEs) both for the RCT sample and for the target population, which we refer to respectively as the \textit{Sample Average Treatment Effect} (SATE) and the \textit{Target Average Treatment Effect} (TATE).
These are defined as the average of the individual additive treatment effects over the RCT sample and over the target population:
\begin{align}
\text{SATE}\equiv\E[Y^1-Y^0|S=1]=\E[Y^1|S=1]-\E[Y^0|S=1],\\
\text{TATE}\equiv\E[Y^1-Y^0|S=0]=\E[Y^1|S=0]-\E[Y^0|S=0].\label{eq:TATE}
\end{align}

Estimation of SATE is straightforward. For simplicity, consider simple randomization, with all RCT participants having the same probability of being assigned treatment.%
\footnote{If the RCT design is complex and treatment probabilities vary across individuals, a minor variation that incorporates inverse-probability-of-treatment weights can be used.}
An unbiased estimate of SATE can be obtained by taking the difference in mean outcome between the treated and control groups, or regressing outcome on treatment adjusting for pre-treatment covariates.
Estimation of TATE, on the other hand, requires adjustment for treatment effect moderators whose distribution differs between the RCT sample and the target population \citep{Olsen2013a,Cole2010}.
The methods for estimating TATE described below assume \textit{conditional sample ignorability for treatment effects} \citep{Kern2016a}: being in the RCT or in the target population sample does not carry any information about treatment effect once we condition on the moderators $Z$.

\subsection{Outcome-model-based TATE estimation}\label{sxn:Z-outcome}

We assume an additive model for the potential outcomes.
With $i$ indexing the individual, the model is
\begin{equation}
\E[Y^t_i]=\beta_0+f_{zt}(Z_i,t)+f_{xz}(X_i,Z_i),~~t=0,1,
\end{equation}
where $f_{zt},f_{xz}$ are functions of the corresponding variables.
For simplicity, we consider the special form $f_{zt}(Z_i,t)=\beta_tt+\beta_{zt}Z_it$, which is perhaps the one most commonly used in practice.
(This form assumes constant moderation effect, as $\beta_{zt}$ does not depend on the level of $Z$.)
The simplified model is
\begin{equation}
\E[Y^t_i]=\beta_0+\beta_tt+\beta_{zt}Z_it+f_{xz}(X_i,Z_i),~~t=0,1.\label{eq:Z-outmod}
\end{equation}
The form of $f_{xz}(X_i,Z_i)$ is not of interest here, but a common practice is to use $\beta_xX_i+\beta_zZ_i$ and perhaps add some complexity such as quadratic or interaction terms.

The treatment effect for individual $i$ is
\begin{equation}
\E[Y^1_i]-\E[Y^0_i]=\beta_t+\beta_{zt}Z_i\label{eq:Z-indvleffect}
\end{equation}
which means
\begin{align}
\text{SATE}&=\beta_t+\beta_{zt}\E[Z|S=1],\label{eq:Z-SATE}\\
\text{TATE}&=\beta_t+\beta_{zt}\E[Z|S=0].\label{eq:Z-TATE-outcome}
\end{align}

The difference between SATE and TATE, $\beta_{zt}\{\E[Z|S=1]-\E[Z|S=0]\}$, is the bias if we generalize the effect estimated in the RCT directly to the target population without adjusting for differences in $Z$. The magnitude of this bias depends on the moderation effect ($\beta_{zt}$) and the difference between the means of the moderator in the two samples ($\{\E[Z|S=1]-\E[Z|S=0]\}$). If either of these is zero, SATE is equivalent to TATE.

When $Z$ is observed in both samples, an estimate for TATE can be obtained using eq. \ref{eq:Z-TATE-outcome}, with $\E[Z|S=0]$ estimated using the target population dataset, and with $\beta_t$ and $\beta_{zt}$ estimated by fitting to the RCT data an outcome model with interaction terms.%
\footnote{An alternative is to estimate the outcome model, predict treatment effects for the individuals in the target population dataset \citep{Kern2016a} using eq. \ref{eq:Z-indvleffect}, and average them. This strategy does not require the constant moderation effect assumption. However, it does not provide for a straightforward sensitivity analysis for an unobserved moderator.}
While eq. \ref{eq:Z-TATE-outcome} does not involve $X$, the accuracy and precision of the estimates of $\beta_t$ and $\beta_{zt}$ require a good estimate of the outcome model. Not only do we need to capture all $Z$ variables, all $X$ variables (or at least all $X$ variables that are correlated with, or interact with, $Z$ variables) should be included and correctly modeled.

Note that we have invoked the \textit{conditional sample ignorability for treatment effects} assumption when equating $\{\beta_t,\beta_{zt}\}$ between equations \ref{eq:Z-SATE} and \ref{eq:Z-TATE-outcome}.
This assumption is violated if we do not observe all the moderators that are differentially distributed between the two samples.
It is also violated if the range of $Z$ in the target population includes segments not covered by the RCT, a violation of the \textit{positivity} assumption \citep{Rosenbaum1983a}; using eq. \ref{eq:Z-TATE-outcome} in this case would result in extrapolation beyond the support of the data.
Positivity is often not a problem with a binary $Z$, but for a continuous $Z$, care needs to be taken to check overlap, and judgment needs to be made about whether extrapolation to any uncovered areas is reasonable.

\subsection{Weighting-based TATE estimation}\label{sxn:Z-weighting}

The idea of this method is to reweight the RCT sample so that it resembles the target population with respect to the distribution of the treatment effect moderators ($Z$) and then to use this weighted RCT sample to estimate TATE.

The weighting procedure involves first stacking the RCT and target population datasets and fitting a model predicting sample membership.
The set of predictors in this model needs to include all the moderators ($Z$ variables); outcome predictors that do not moderate treatment effect ($X$ variables) do not need to be included.
To determine which pre-treatment covariates are moderators requires a prior step of detecting them through modeling the outcome.
There may be times when it is hard to know whether a variable is a moderator (e.g., its interaction term with treatment has a substantial but statistically non-significant coefficient), in which case it is preferable to treat it as a moderator and include it in the sample membership model. 
For the same reason (or to avoid having to model the outcome), one may also include a broader set of variables in this model, regardless of whether they may be moderators ($Z$) or not ($X$).

The fitted sample membership model is used to compute the predicted odds of being in the target population sample for the RCT participants. These odds are then used to reweight the RCT sample. As a result, the weighted RCT sample better resembles the target population with respect to the distribution of the variables used in the sample membership model.
%\footnote{This weighting scheme weights the whole RCT sample to the target population sample. An alternative is to weight each arm of the RCT to the target population sample.}
This strategy of weighting the RCT sample to the target population sample has been described by \citet{Kern2016a} and \citet{Cole2010};%
\footnote{The weights are the same in \citet{Kern2016a}, but slightly different in \citet{Cole2010} because in the latter case the RCT sample was a subsample of the target population dataset. We will return to this in the Discussion section.}
here we emphasize the distinction between moderators and other covariates, as the purpose of the weighting is to adjust for the diffential distribution of the moderators. Whether the weighting succeeds in doing this should be checked.

The weighted RCT sample is used to estimate an average treatment effect, which is taken as the estimated TATE.
A simple estimator for TATE is the difference between the weighted means of the outcome in the RCT's treated and control groups.
Another option is to fit a weighted regression model that controls for $Z$ and $X$ variables (but not their interaction terms with $T$), and estimate  TATE with the coefficient of $T$.%
%\footnote{An alternative to this weight-and-estimate approach is the approach used by \citet{Kern2016a}: fitting a flexible outcome model allowing interactions to the RCT sample to get effect estimates for different levels of $Z$, and averaging these over the target population distribution of $Z$. We will comment on this approach in the Discussion section.}

\section{Sensitivity analysis for a moderator that is observed in the RCT but not in the target population sample}\label{sxn:V}

We continue using $Z$ to denote moderators observed in both samples, and use $V$ to denote a moderator observed in the RCT but not in the target population sample.
(We hereafter refer to the current case as the $V$ case, to differentiate it with the case %of a moderator not observed even in the RCT, 
to be addressed in Section \ref{sxn:U}.)
In this case, although TATE cannot be estimated in a way that adjusts for all of the moderators, we can conduct sensitivity analysis to assess how TATE estimates would change based on what we assume about the distribution of $V$ in the target population.
Here we present several sensitivity analysis methods, and report on two simulation studies that compare some of these methods to one another.

\subsection{Three sensivity analysis strategies}

The methods described below are based on an outcome model, full weighting adjustment, and partial weighting combined with an outcome model.

\subsubsection{Outcome-model-based sensitivity analysis}\label{sxn:V-outcome}

We rewrite the potential outcomes model, separating $Z$ and $V$:
\begin{equation}
\E[Y^t_i]=\beta_0+\beta_tt+\beta_{zt}Z_it+\beta_{vt}V_it+f_{xzv}(X_i,Z_i,V_i).\label{eq:V-outmod}
\end{equation}
For simplicity, this model makes an additional assumption (compared to the model in eq. \ref{eq:Z-outmod}) that there is no three-way interaction of the treatment with both $Z$ and $V$.
Based on this model, the formula for TATE is
\begin{align}
\text{TATE}&=\beta_t+\beta_{zt}\E[Z|S=0]+\beta_{vt}{\color{red}{\E[V|S=0]}},\label{eq:V-TATE-outcome}
\end{align}
where $\beta_t,\beta_{zt},\beta_{vt},\E[Z|S=0]$ can be estimated from data, whereas $\color{red}{\E[V|S=0]}$ cannot.
We will refer to the latter as an `unknown' parameter, which is a slight abuse of terminology because the true values of all these parameters, $\beta_t,\beta_{zt},\beta_{vt},\E[Z|S=0]$ and $\color{red}{\E[V|S=0]}$, are not known.
By `unknown' here, we mean that one cannot learn about this parameter from data, while one can learn about the other parameters from data.

The simple formula in eq. \ref{eq:V-TATE-outcome} results from the no three-way interaction assumption.
Without such assumption, the potential outcomes model would have an additional term, $\beta_{zvt}Z_iV_it$, and the formula for TATE would include $\beta_{zvt}\E[ZV|S=0]$, with the unknown $\E[ZV|S=0]$ being more complex to consider than simply $\E[V|S=0]$.

To conduct the sensitivity analysis, first we need to estimate the estimable quantities.
$\E[Z|S=0]$ is estimated using target population data.
Assuming sample ignorability for treatment effects conditional on $Z,V$, we estimate $\beta_t,\beta_{zt},\beta_{vt}$ using the RCT data; this involves estimating the outcome model with interaction terms (eq. \ref{eq:V-outmod}) in the same manner as discussed in section \ref{sxn:Z-outcome}, and extracting the estimated values and variance-covariance matrix of $\beta_t,\beta_{zt},\beta_{vt}$.

We then specify a plausible range for the unknown $\color{red}{\E[V|S=0]}$ (mean $V$ in the target population).
In doing this, it is important to check if the range of $V$ being considered has good overlap with the RCT sample.

A range for the TATE point estimate is computed by plugging the point estimates of $\beta_t,\beta_{zt},\beta_{vt},\E[Z|S=0]$ and the specified range of $\color{red}{\E[V|S=0]}$ into eq. \ref{eq:V-TATE-outcome}.

A confidence band to accompany this TATE range can be obtained.
For each value of $\color{red}{\E[V|S=0]}$ in the specified range, the variance-covariance matrix of the estimated $\beta_t,\beta_{zt},\beta_{vt}$ can be used to obtain a confidence interval for TATE.
If the uncertainty in the estimated $\E[Z|S=0]$ is non-negligible, it can be incorporated by using the confidence limits of $\E[Z|S=0]$ (rather than its point estimate) in the construction of such confidence intervals.

\subsubsection{Weighting-based sensitivity analysis}\label{sxn:V-weighting}

Ideally, had $V$ been available from both samples, we would be able to estimate TATE using RCT data, weighting the individuals by their odds of being in the target population sample conditional on $Z,V$ (as described in section \ref{sxn:Z-weighting}).
While such weights cannot be estimated when $V$ is not observed in the target population, they can be reexpressed, using Bayes' rule, as
\begin{align}
W_i
&=\frac{\text{P}(S=0|Z_i,V_i)}{\text{P}(S=1|Z_i,V_i)}
\nonumber\\
&=\frac{\text{P}(S=0,Z=Z_i,V=V_i)/\text{P}(Z=Z_i,V=V_i)}{\text{P}(S=1,Z=Z_i,V=V_i)/\text{P}(Z=Z_i,V=V_i)}\nonumber\\
&=\frac{\text{P}(S=0,Z=Z_i,V=V_i)}{\text{P}(S=1,Z=Z_i,V=V_i)}\nonumber\\
&=\frac{\text{P}(Z=Z_i)\text{P}(S=0|Z_i)\text{P}(V=V_i|S=0,Z_i)}{\text{P}(Z=Z_i)\text{P}(S=1|Z_i)\text{P}(V=V_i|S=1,Z_i)}\nonumber\\
&=\frac{\text{P}(S=0|Z_i)}{\text{P}(S=1|Z_i)}\cdot\frac{\color{red}{\text{P}(V=V_i|S=0,Z_i)}}{\text{P}(V=V_i|S=1,Z_i)}.\label{eq:V-fullweights}
\end{align}
Each weight is thus a product of two components: (1) the odds of being in the target population sample conditional on $Z$ but not $V$, and (2) a ratio of the probability density/mass of $V=V_i$ in the $Z=Z_i$ stratum comparing the target population sample to the RCT sample.%
\footnote{This ratio is of the same form as a ratio used elsewhere in weighting to control confounding in causal mediation analysis \citep{Hong2010}.}

In this formula of the weights (eq. \ref{eq:V-fullweights}), the first component can readily be estimated from data; the denominator of the second component can also be estimated. The numerator of the second component, $\color{red}{\text{P}(V=V_i|S=0,Z_i)}$, is unknown.
This suggests that a sensitivity analysis can be conducted by specifying a plausible range for the unknown distribution of $V$ given $Z$ in the target population, $\color{red}{\text{P}(V|S=0,Z)}$, and for each distribution in this range, constructing weights and estimating TATE using the reweighted RCT sample.
TATE can be estimated using either the difference in weighted mean outcome between the treated and control conditions, or using regression of the outcome on treatment and covariates.

The challenge is how to estimate $\text{P}(V|S=1,Z)$ and how to specify plausible ranges for $\color{red}{\text{P}(V|S=0,Z)}$.
Both these tasks are complicated and results are prone to misspecification bias when $V$ or $Z$ or both are of any form but binary.
We hereby limit the consideration of this method to the case where $V$ and $Z$ are binary.
With one binary $Z$ and one binary $V$, there are only four unique weights:
\begin{align*}
{W_i}_{|V_i=1,Z_i=1} & =\frac{\text{P}(S=0|Z=1)}{\text{P}(S=1|Z=1)}\cdot\frac{\color{red}{\text{P}(V=1|S=0,Z=1)}}{\text{P}(V=1|S=1,Z=1)},\nonumber\\
{W_i}_{|V_i=0,Z_i=1} & =\frac{\text{P}(S=0|Z=1)}{\text{P}(S=1|Z=1)}\cdot\frac{1-\color{red}{\text{P}(V=1|S=0,Z=1)}}{1-\text{P}(V=1|S=1,Z=1)},\nonumber\\
{W_i}_{|V_i=1,Z_i=0} & =\frac{\text{P}(S=0|Z=0)}{\text{P}(S=1|Z=0)}\cdot\frac{\color{red}{\text{P}(V=1|S=0,Z=0)}}{\text{P}(V=1|S=1,Z=0)},\nonumber\\
{W_i}_{|V_i=0,Z_i=0} & =\frac{\text{P}(S=0|Z=0)}{\text{P}(S=1|Z=0)}\cdot\frac{1-\color{red}{\text{P}(V=1|S=0,Z=0)}}{1-\text{P}(V=1|S=1,Z=0)}.
\end{align*}
The denominators in the second component of these weights are easily estimated. For the numerators, we need to specify ranges for two probabilities: $\color{red}{\text{P}(V=1|S=0,Z=1)}$ and $\color{red}{\text{P}(V=1|S=0,Z=0)}$, the prevalence of $V=1$ in the target population given $Z=1$ and $Z=0$.

\subsubsection{Weighted-outcome-model-based sensitivity analysis}\label{sxn:V-wtdoutcome}

While the full weighting strategy is hard to implement in the context of sensitivity analysis, a partial weighting version combined with an outcome model lends itself well to sensitivity analysis.
The idea is to use weighting to adjust for known differences between the two samples (here differences in the distribution of $Z$) and then to use an outcome model to do sensitivity analysis on unknown quantities ($V$ in the target population).
First, we weight the RCT sample by the individuals' odds of being in the target population sample conditional on $Z$ only, $\frac{\text{P}(S=0|Z_i)}{\text{P}(S=1|Z_i)}$. We then use this weighted RCT sample to estimate the outcome model (of the form in eq. \ref{eq:V-outmod}). We use the estimates of $\beta_t,\beta_{zt},\beta_{vt}$ and their variance-covariance matrix from this model as inputs for estimating TATE in the same manner as in the non-weighted \textit{outcome-model-based} method (based on TATE formula eq. \ref{eq:V-TATE-outcome}).

The difference between this method and the first method is that the weighting makes the distribution of $Z$ in the RCT sample more similar to that in the target population, and thereby helps adjust for the discrepancy in average treatment effect due to effect moderation by $Z$. This may be helpful in the case the $Z$ part of the outcome model is misspecified, which we investigate in a simulation study reported in section \ref{sxn:V-simulation2}.

\subsection{Simulation study comparing the outcome-model-based and weighted-outcome-model-based sensitivity analyses}\label{sxn:V-simulation2}

We investigate how well these two methods perform relative to each other in recovering the true TATE, in situations where the outcome model is correctly or incorrectly specified.
When the outcome model is correctly specified, we expect that both methods are unbiased.
When the $Z$ part of the outcome model is misspecified, we expect that the \textit{weighted-outcome-model-based} method is less biased.
When the $V$ part of the outcome model is misspecified, we expect that the same method helps reduce bias due to this misspecification if $Z$ and $V$ are positively correlated and influence treatment effect in the same direction.

\subsubsection{Data generation} We consider situations with one $X$, one $Z$ and one $V$. $X$ is a standard normal random variable. $Z$ and $V$ are first generated as multivariate normal with correlations ranging from 0 to $\pm .5$, and then each is either kept in continuous form or dichotomized. When either $Z$ or $V$ is binary, its prevalence is .25 in the RCT sample and .5 in the target population. When either $Z$ or $V$ is continuous, it has mean 0 in the RCT and .5 in the target population, and variance 1 in both.

In the RCT, $T$ is randomly assigned to 0 and 1 with probability 0.5. With regards to the outcome, for the continuous $Z$ and $V$ combination, we use a base model with $Z$ and $V$ as moderators, plus six other models, each with one additional moderator from among $Z^2$, $V^2$ or $ZV$, whose moderation effect is either positive or negative.

\begin{align*}
\begin{matrix*}[l]
\text{A.}  & Y=X+T+Z+V+ZT+VT+\epsilon_Y\\
\text{B1.} & Y=X+T+Z+V+ZT+VT+Z^2T+\epsilon_Y\\
\text{B2.} & Y=X+T+Z+V+ZT+VT-Z^2T+\epsilon_Y\\
\text{C1.} & Y=X+T+Z+V+ZT+VT+V^2T+\epsilon_Y\\
\text{C2.} & Y=X+T+Z+V+ZT+VT-V^2T+\epsilon_Y\\
\text{D1.} & Y=X+T+Z+V+ZT+VT+ZVT+\epsilon_Y\\
\text{D2.} & Y=X+T+Z+V+ZT+VT-ZVT+\epsilon_Y
\end{matrix*}~~~,~~
\epsilon_Y\sim \text{N}(0,4)
\end{align*}
For the continuous $Z$ and binary $V$ combination, we use models A, B1-2 and D1-2. For the binary $Z$ and continuous $V$ combination, we use A, C1-2 and D1-2. For the binary $Z$ and $V$ combination, we use A and D1-2.

For each scenario (combining $Z$ and $V$ types and outcome model), 100,000 $n$=400 RCT samples and $n$=5000 target population samples are generated.

\subsubsection{Methods implementation}

\paragraph{Outcome models used in the sensitivity analyses}

For scenarios with the true outcome model A, we implement the \textit{outcome-model-based} and \textit{weighted-outcome-model-based} sensitivity analyses using the correctly specified outcome model.
For the other scenarios, we implement these methods using the correct model as well as the misspecified model leaving out the third moderator ($Z^2$, $V^2$ or $ZV$).
We choose to consider this misspecified model because it is simple and perhaps most often used.
In practice, detection of moderation effects using regression is often an exploratory analysis trying out interaction terms of different covariates with treatment.
More complex interaction terms are less often considered, and even if they are, the power to detect them is limited.

\paragraph{Weighting details}

With the \textit{weighted-outcome-model-based} method, the weighting is partial, adjusting for $Z$ but not $V$. We compute the weights are based on a logistic sample membership model with $Z$ as the predictor. With a continuous $Z$, to allow flexible modeling of sample membership probability, we use natural splines with nine knots.

\begin{figure}
\caption{Bias of \textit{outcome-model-based} and \textit{weighted-outcome-model-based} sensitivity analyses using correct and misspecified outcome models.}
\centering
\includegraphics[width=\textwidth]{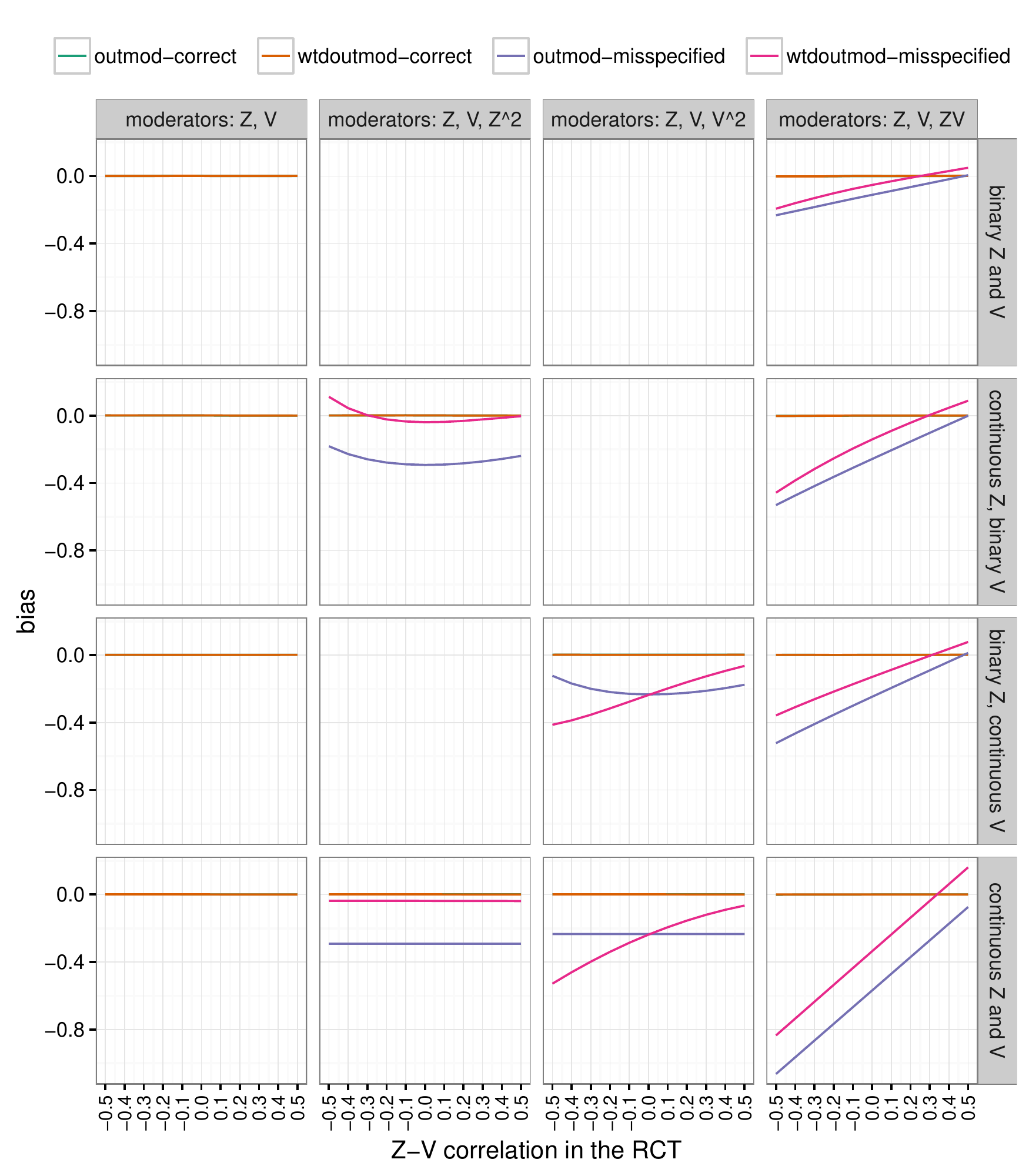}
\caption*{\footnotesize{Notes: `outmod' = outcome-model-based; `wtdoutmod' = weighted-outcome-model-based. For both methods, the same outcome models are used. In scenarios with only two moderators ($Z$ and $V$), only the correct model is used. In scenarios with a third moderator ($Z^2$, $V^2$ or $ZV$), the correct model and the misspecified model that excludes the third moderator are used. In all plots, the green curve (the outcome-model-based method using the correct model) lies underneath the brown curve (the weighted-outcome-model-based method also with the correct model) and is thus not visible. In these scenarios the moderation effects of $Z^2$, $V^2$ or $ZV$ are positive. Results from scenarios where they are negative are mirror images of these plots across the horizontal zero line, i.e., the sign of bias is flipped.}}
\label{figure:bias}
\end{figure}

\subsubsection{Simulation results}

The main results to report concern the bias or lack thereof of the two methods under investigation when using correct or misspecified outcome models.
Since the true ATE in a particular target population sample may differ from the true TATE as set by the simulation parameters, we use the true ATE for a particular target population sample as the goal of inference for each simulation iteration.
This helps avoid noise due to sampling variability and focuses on the bias itself.

Figure \ref{figure:bias} presents the bias of these two methods using correct and misspecified outcome models. Across all scenarios, when the correct outcome model is used, both methods are unbiased. When a misspecified model is used, both methods are biased. When the model is correctly specified with respect to $V$ but misspecified with respect to $Z$ (in scenarios with $Z^2$ as the third moderator -- column 2 in the Figure), as expected, the weighted-outcome-model-based method, which uses weighting to adjust for $Z$, is less biased than the outcome-model-based method.

When the model is correctly specified with respect to $Z$ but misspecified with respect to $V$ (scenarios with $V^2$ as the third moderator -- column 3), the two methods are similarly biased if $Z$ and $V$ are uncorrelated. When $Z$ and $V$ are positively correlated, the weighted-outcome-model-based method becomes less biased, because the weighting adjustment for $Z$ provides some adjustment for $V$. When $Z$ and $V$ are negatively correlated, the contrary is true, with the weighted-outcome-model-based method being more biased.

When the model is misspecified with respect to both $Z$ and $V$ (scenarios with $ZV$ as the third moderator -- column 4), both methods are biased. For most of the range of $Z$-$V$ correlation considered, the weighted-outcome-model-based method is less biased. At a certain point when the correlation is high enough the side-effect adjustment for $V$ that results from weighting adjustment for $Z$ pulls the TATE estimate to zero bias and then past zero to bias of opposite sign.

These results confirm our hypotheses that (i) when the outcome model is misspecified with respect to $Z$, the weighted-outcome-model-based method is less biased than the outcome-model-based method; and that (ii) when the outcome model is misspecified with respect to $V$, and $Z$ and $V$ are positively correlated and influence treatment effect in the same direction, the weighted-outcome-model-based method tends to be less biased.

\section{Sensitivity analysis for effect moderation that is completely unobserved}\label{sxn:U}

In this situation, there may be known moderators ($Z$) that we observe in both the RCT and the target population sample.
We are concerned, however, that there may be additional effect moderation by factors that are unobserved even in the RCT that may be differentially distributed between the RCT sample and the target population.
While in the previous situation we deal with a specific variable $V$ (specific because it is observed in the RCT), here we think of a generic variable denoted by $U$ that is unknown and unobserved.
At this point in our conceptualization, $U$ can potentially be an actual characteristic that is unobserved, or it can be a quantity that combines multiple effect moderation factors.
Our purpose is to determine whether sensitivity analyses for the $V$ case can be adapted for the current case (which we refer to as the $U$ case).

The full-weighting method requires $U$ to be observed in the RCT, so cannot be used here.
The outcome-model-based methods, with a formula for TATE that includes the $\beta_t$ term (like the formula in eq. \ref{eq:V-TATE-outcome}, except replacing $V$ with $U$), also cannot be used because the estimation of $\beta_t$ depends on the variables interacting with treatment, including $U$.
However, a small modification results in methods that work for a special definition of $U$.

\subsection{Two sensitivity analyses for the $U$ case}

\subsubsection{Bias-formula-based sensitivity analysis}

Assume the linear potential outcomes model
\begin{equation}
\E[Y^t_i]=\beta_0+\beta_tt+\beta_{zt}Z_it+\beta_{ut}U_it+f_{xzu}(X_i,Z_i,U_i)\label{eq:U-outmod}
\end{equation}
similar to the model for the $V$ case, also with no three-way interaction with treatment.
With the assumption of sample ignorability for treatment effects conditional on $Z,U$, we have both
\begin{align}
\text{SATE}&=\beta_t+\beta_{zt}\E[Z|S=0]+{\color{red}{\beta_{ut}\E[U|S=1]}},~\text{and}\label{eq:U-SATE-outcome}\\
\text{TATE}&=\beta_t+\beta_{zt}\E[Z|S=0]+{\color{red}{\beta_{ut}\E[U|S=0]}}.\label{eq:U-TATE-outcome}
\end{align}
Note that this assumption requires $U$ to capture all effect moderating forces other than $Z$, thus narrowing the definition of $U$.
Eq. \ref{eq:U-SATE-outcome} and eq. \ref{eq:U-TATE-outcome} imply that 
\begin{align}
\text{TATE}=\text{SATE}&+\beta_{zt}\{\E[Z|S=0]-\E[Z|S=1]\}+\nonumber\\
&+{\color{red}{\beta_{ut}\{\E[U|S=0]-\E[U|S=1]\}}}.\label{eq:U-TATE-bias1}
\end{align}
On the right hand-side of eq. \ref{eq:U-TATE-bias1}, SATE can be estimated unbiasedly as the difference in mean outcome between the two treatment conditions in the RCT.
To use eq. \ref{eq:U-TATE-bias1} for sensitivity analysis, we need an unbiased estimate of $\beta_{zt}$.
Like in the $V$ case, the model used to estimate $\beta_{zt}$ needs to include all $X$ variables that are correlated, or interact, with $Z$.
Leaving $U$ out of the model, however, generally leads to bias in the estimated $\beta_{zt}$ (as well as other coefficients).
The only situation where omitting $U$ would not bias $\beta_{zt}$ is when $U$ is independent of $Z$.
This requires further refining the definition of $U$ to a quantity that combines all the unobserved moderating factors after `regressing out' $Z$.
We call this variable the \textit{remaining composite moderator after accounting for $Z$}, and denote it by $U_{(z)}$.%
\footnote{This consideration of $U_{(z)}$ independent of all observed moderators $Z$ parallels the convention of evaluating treatment effect sensitiveness to an unobserved confounder independent of observed confounders \citep{Rosenbaum1983}.}

With $U_{(z)}$
so defined,
%being the \textit{remaining composite moderator after accounting for $Z$},
we can estimate $\beta_{zt}$ and use
\begin{align}
\text{TATE}=\text{SATE}&+\beta_{zt}\{\E[Z|S=0]-\E[Z|S=1]\}+\nonumber\\
&+{\color{red}{\beta_{ut}\{\E[U_{(z)}|S=0]-\E[U_{(z)}|S=1]\}}}\label{eq:U-TATE-bias2}
\end{align}
for sensitivity analysis. 
By varying $\color{red}{\beta_{ut}\{\E[U_{(z)}|S=0]-\E[U_{(z)}|S=1]\}}$, we get a range for the point estimate of  TATE.
We will address how to specify ranges for such an unknown quantity after discussing the \textit{weighting-plus-bias-formula-based} method.

\subsubsection{Weighting-plus-bias-formula-based sensitivity analysis}

With this approach, we have the option of weighting the RCT sample to adjust for $Z$ and conducting a bias-formula-based sensitivity analysis for a $U$ that is independent of $Z$ (the remaining composite moderator after accounting for $Z$).
Yet it is plausible that $X$ variables may carry some (even if limited) information about unobserved moderators---they may be correlated with unobserved moderators but the correlations are small so $X$ do not appear to be moderators themselves.
We therefore propose adjusting for both $X$ and $Z$ through weighting and then conducting sensitivity analysis for a $U$ independent of $X,Z$ (the \textit{remaining composite moderator after accounting for $X,Z$}).
We denote this variable by $U_{(xz)}$.

We weight the individuals in the RCT by their odds of being in the target population sample conditional on $X,Z$.
The weighted RCT sample now better resembles the target population sample with respect to the distribution of $X,Z$.
On the other hand, it resembles the unweighted RCT sample with respect to the distribution of $U$, because $U$ is independent of $X,Z$.
We call the ATE estimated from this weighted sample the $X$\textit{-and-}$Z$\textit{-adjusted} ATE (xzATE).
Based on the potential outcomes model,
\begin{align}
\text{xzATE}
&=\beta_t+\beta_{zt}\E[Z|S=1,\text{xz-wtd}]+\beta_{ut}\E[U_{(xz)}|S=1]\\
&\approx\beta_t+\beta_{zt}\E[Z|S=0]+\beta_{ut}\E[U_{(xz)}|S=1]
\end{align}
(where `xz-wtd' stands for `weighted to adjust for $X,Z$'). This means
\begin{align}
\text{TATE}=\text{xzATE}&+\beta_{zt}\{\E[Z|S=0]-\E[Z|S=1,\text{xz-wtd}]\}+\nonumber\\
&+{\color{red}{\beta_{ut}\{\E[U_{(xz)}|S=0]-\E[U_{(xz)}|S=1]\}}}.\label{eq:U-TATE-wtdbias1}
\end{align}
and
\begin{equation}
\text{TATE}\approx\text{xzATE}+{\color{red}{\beta_{ut}\{\E[U_{(xz)}|S=0]-\E[U_{(xz)}|S=1]\}}}.\label{eq:U-TATE-wtdbias2}
\end{equation}

An unbiased estimate for xzATE is the difference between the weighted means of the outcome in the two treatment conditions in the RCT.
If the weighting succeeds in equating the means of $Z$ between the RCT and target population datasets, eq. \ref{eq:U-TATE-wtdbias2} can be used for sensitivity analysis.
If the weighting reduces the distance between these means but not to zero, eq. \ref{eq:U-TATE-wtdbias1} can be used.
If eq. \ref{eq:U-TATE-wtdbias1} is used, we get a range for TATE point estimates corresponding to the plausible range specified for the unknown $\color{red}{\beta_{ut}\{\E[U_{(xz)}|S=0]-\E[U_{(xz)}|S=1]\}}$.
If eq. \ref{eq:U-TATE-wtdbias2} is used, in addition to the point estimate range, we also get confidence limits for TATE.

\paragraph{Plausible range specification for sensitivity parameters}

Both of the bias-formula-based methods require specifying some plausible range for the unknown $\color{red}{\beta_{ut}\{\E[U|S=0]-\E[U|S=1]\}}$ where $U$ is either $U_{(z)}$ or $U_{(xz)}$.
This quantity can be considered the combination of two sensitivity parameters: one representing the moderation effect ($\color{red}{\beta_{ut}}$) and the other representing the association between $U$ and sample membership (the difference in mean $U$ between the two samples, $\color{red}{\E[U|S=0]-\E[U|S=1]}$).
As the remaining composite moderator (combining potentially multiple moderating factors), the most appropriate form for $U$ to take is perhaps the form of a continuous variable.
We propose using a standardized metric here, so the difference in mean $U$ between the two samples is in standard deviation units, and $\beta_{ut}$ is the change in treatment effect associated with one standard deviation difference in $U$.

\paragraph{Alternative conceptualization of $U$ as a natural variable}

The definition of $U$ as a composite variable---representing the remaining effect moderation factors after accounting for $Z$ or for $X,Z$---requires some degree of abstraction away from real world quantities.
It may be common, however, for scientists to think in more concrete terms, asking whether there may exist an unobserved natural variable (as opposed to a composite variable) that moderates treatment effect and that is differentially distributed between the target population and the RCT sample.
It is important to note that this is a special-case interpretation of $U$, and that it requires that (1) this unobserved natural variable is the only unobserved moderator, and that it is either (2a) independent of $X,Z$ (if using the weighting-plus-bias-formula-based method and weighting to adjust for $X,Z$), or (2b) independent of $Z$ (if using the bias-formula-based method or if using the weighting-plus-bias-formula-based method but weighting to adjust for $Z$ only).
In this special case where $U$ is a natural variable, it can be of any form, e.g., continuous, dichotomous, polytomous, etc.

\section{Real data application}\label{sxn:application}

We consider a smoking cessation RCT for drug and/or alcohol-dependent adults \citep{Reid2008}, known as CTN9 in NIDA's Clinical Trial Network's repository of substance use treatment RCTs.
Participants (n=225) were adult cigarette smokers who at baseline smoked at least 10 cigarettes per day, recruited from among people who attended outpatient community-based treatment programs for opiate, cocaine and alcohol dependence. They were randomly assigned in a 2:1 ratio to receive either smoking cessation treatment or no such treatment.
Smoking cessation treatment consisted of one week of group counseling before the target quit date and eight weeks of group counseling plus transdermal nicotine patch treatment (21 mg per day for weeks 1 to 6 and 14 mg per day for weeks 7 and 8) after the target quit date.
We retain 200 participants in analysis (including 65 treated and 135 controls), excluding 18 with no outcome data, and then an additional seven who were either in a controlled environment, or Asian, Pacific Islanders, and Native Americans, since generalizing from such small numbers would be inadvisable, given the plausibility of these categories as effect moderators.

Given NIDA's interest in using NIDA-supported RCTs to generate evidence relevant to practice, we define the target population to be adults in the US who seek treatment for alcohol/substance use disorders who also smoke at least 10 cigarettes per day.
To represent this target population, we use a subset of the 2014 National Survey on Drug Use and Health (\mbox{NSDUH}), a representative sample of the US population excluding homeless persons outside shelters, active duty personnel, and those in controlled environments.
Out of 55,271 NSDUH respondents, 2,751 were adults (aged 18 and older) who reported having ever sought treatment for substance abuse, excluding Asians, Pacific Islanders, and Native Americans.
Of those, 934 reported smoking at least 10 cigarettes a day on average, comprising our target population sample.

Table \ref{table:1} summarizes baseline characteristics of the RCT sample (including demographics, education, employment, baseline smoking, cigarette addiction severity score, number of past quit attempts, years smoking, reasons for quitting, and primary substance of abuse -- see \citet{Reid2008} for detailed description) and the same variables (if available) from the target population sample.
The two samples differ in all the characteristics observed in both: the RCT sample has larger proportions of Hispanic, African-American and female participants, is older, and smokes more on average.

\begin{table}
\caption{Baseline characteristics of the RCT (n=200) and target population (n=934) samples}
\centering
%\resizebox{\textwidth}{!}{ 
\begin{tabular}{lrrrrr}
& & && \multicolumn{2}{c}{Target}\\
& \multicolumn{2}{c}{RCT} && \multicolumn{2}{c}{population}\\ \hline
\textbf{Demographics}\\
Male gender: number (\%) & 105 & (52.5) && 587 & (62.8)\\
Race/ethnicity: number (\%)\\
~~White & 73 & (36.5) && 764 & (81.8)\\
~~African-American & 51 & (25.5) && 67 & (7.2)\\
~~Hispanic & 61 & (30.5) && 58 & (6.2)\\
~~Multiple & 15 & (7.5) && 45 & (4.8)\\
Age in years: mean (SD) & 42.3 & (9.6) && 36.9 & (12.3)\\
Years of education: mean (SD) & 11.5 & (2.1)\\
Employment: number (\%)\\
~~Full-time & 49 & (24.5)\\
~~Part-time or student & 25 & (12.5)\\
~~Retired or unemployed & 126 & (63.0)\\ \hline
\textbf{Primary substance abuse}\\
Primary substance of abuse: number (\%)\\
~~Opiates & 113 & (56.5)\\
~~Cocaine & 39 & (19.5)\\
~~Alcohol/other & 48 & (24.0)\\
Severity of primary substance abuse: mean (SD) & 0.76 & (1.04)\\ \hline
\textbf{Cigarette smoking and addiction}\\
Daily number of cigarettes: mean (SD) & 21.2 & (11.3) && 17.6 & (8.4)\\
Number of smoking years: mean (SD) & 26.4 & (9.9)\\
Number of quit attempts: mean (SD) & 5.2 & (12.6)\\
Urine cotinine: mean (SD) & 1209 & (667)\\
Addiction severity score: mean (SD) & 4.05 & (0.78)\\
Withdrawal scale: mean (SD) & 1.68 & (0.98)\\ \hline
SD = standard deviation.
\end{tabular}
%}
\label{table:1}
\end{table}

Reid and colleagues analyzed the RCT data using a longitudinal model with the daily numbers of cigarettes smoked (collected once a week during active treatment and at three and sixteen weeks after treatment) as repeated outcome measures.
They found a significant reduction in the number of cigarettes smoked per day in the treatment group.
In our analysis, we use the mean number of cigarettes smoked per day over the eight weeks after the target quit date as the outcome variable.
This is justifiable since after the target quit date, the number of cigarettes smoked by the treatment group declined and stayed at about the same level throughout the end of treatment.

To get an estimate of SATE, we fit a linear model for the outcome that includes treatment condition and all the baseline covariates in Table \ref{table:1}.
Consistent with findings by Reid and colleagues, we find a significant average decrease of ten cigarettes a day as a result of the treatment.
We then explore treatment effect heterogeneity, considering models with the same variables (as main effects) plus covariate-treatment interactions.
For model selection, we use stepwise regression with forward selection and backward elimination, minimizing the Akaike information criterion.
The model selected, presented in Table \ref{table:2}, includes interaction terms of treatment with African-American race cateogory, baseline number of cigarettes per day, and with baseline cigarette addiction severity. Specifically, African-American participants and participants who smoked a larger number of cigarettes per day at baseline experienced a larger reduction, and those with higher baseline addiction score experienced a smaller reduction, in cigarettes smoked per day.

\begin{table}[h!]
\caption{Treatment effects from the outcome model with interaction terms}
\centering
%\resizebox{\textwidth}{!}{ 
\begin{tabular}{lrr}
Variable & Coefficient & 95\% confidence interval\\ \hline
Treatment & $-6.15$ & $(-14.37,-2.06)$\\
Treatment X African-American & $-4.09$ & $(-8.09,-0.09)$\\
Treatment X baseline daily number of cigarettes & $-0.60$ & $(-0.76,-0.43)$\\
Treatment X baseline cigarette addiction severity & $2.37$ & $(0.33,4.41)$ \\ \hline
\multicolumn{3}{l}{\footnotesize This model also includes all covariates in Table \ref{table:1} as main effects.}
\end{tabular}
%}
\label{table:2}
\end{table}

\begin{table}[h!]
\caption{Treatment effects from the weighted outcome model with interaction terms}
\centering
%\resizebox{\textwidth}{!}{ 
\begin{tabular}{lrr}
Variable & Coefficient & 95\% confidence interval\\ \hline
Treatment & $-3.70$ & $(-17.13,9.72)$\\
Treatment X African-American & $-4.49$ & $(-8.11,-0.87)$\\
Treatment X baseline daily number of cigarettes & $-0.55$ & $(-0.82,-0.27)$\\
Treatment X baseline cigarette addiction severity & $1.77$ & $(-1.77,5.30)$ \\ \hline
\multicolumn{3}{l}{\footnotesize This model was fit to RCT data weighted to adjust for the differential distribution of}\\
\multicolumn{3}{l}{\footnotesize race, gender, age and baseline daily number of cigarettes between the two samples. The}\\
\multicolumn{3}{l}{\footnotesize model includes all covariates in Table \ref{table:1} as main effects.}
\end{tabular}
%}
\label{table:3}
\end{table}

\begin{figure}
\caption{Results of the \textit{outcome-model-based} and \textit{weighted-outcome-model-based} sensitivity analyses with the data application.}
\centering
\includegraphics[width=.65\textwidth]{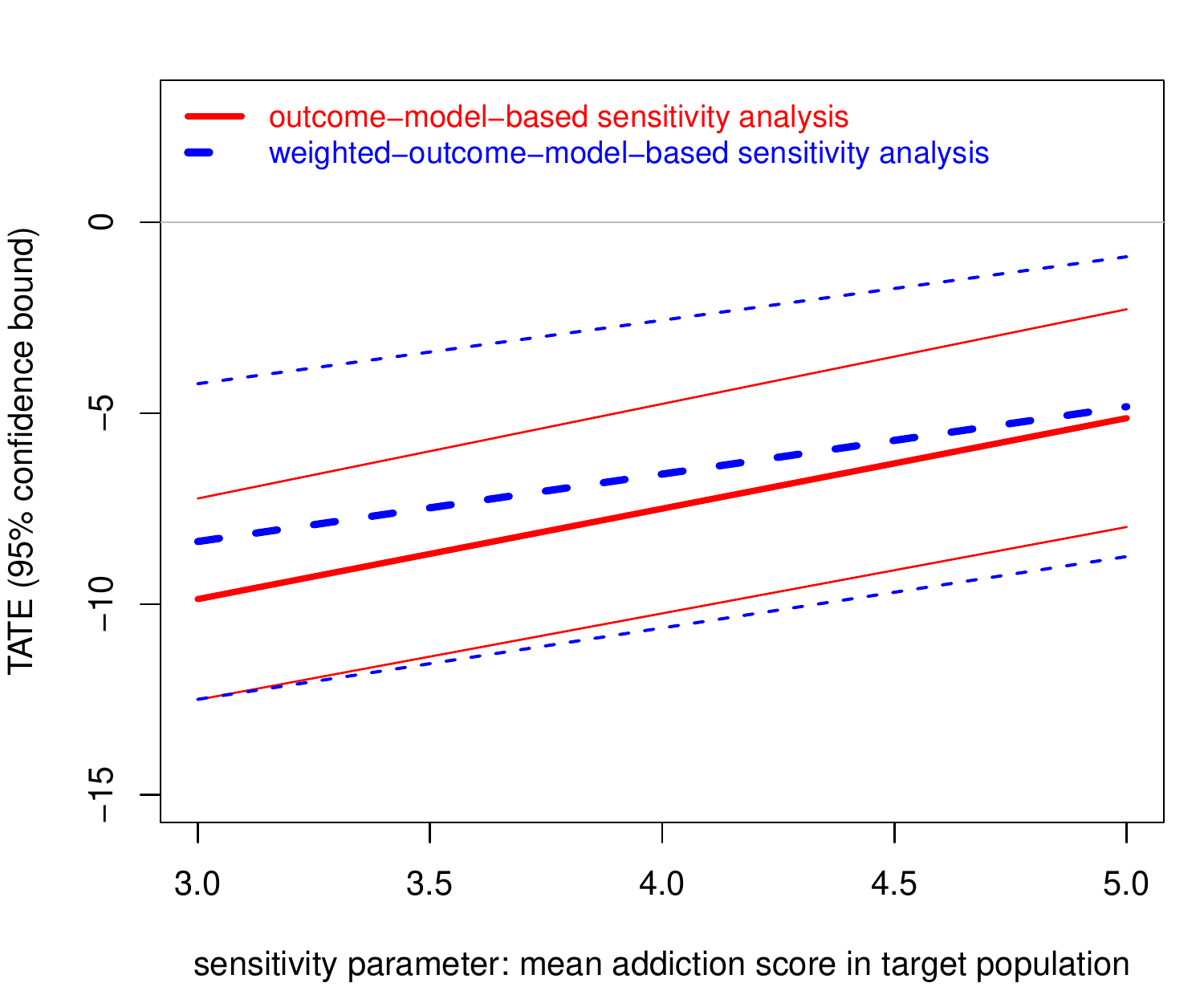}
\label{figure:dataexample}
\end{figure}

A key problem with trying to genearlize the RCT results to the target population is that the baseline cigarette addiction severity score is a treatment effect moderator, but its target population distribution is unknown, because this variable is not available from the NSDUH.  The methods described in this paper can be used to generate TATE estimates accounting for possible distributions of cigarette addiction severity in the target population. In the RCT, the mean of this variable is 4.05 (on a 1 to 5 scale). We assume that the mean cigarette addiction severity score in the target population is somewhere between 3 and 5, and let this sensitivity parameter vary over this range. Applying the outcome-model-based method described in section \ref{sxn:V-outcome}, we have TATE ranging from $-9.87$, 95\% CI $(-12.51,-7.23)$ (corresponding to mean baseline addiction score 3) to $-5.13$, 95\% CI $(-7.98,-2.28)$ (corresponding to mean baseline addiction score 5).%a range and 95\% confidence bound for TATE presented in the top plot in Figure \ref{figure:dataexample}.

For the weighted-outcome-model-based method, the first step is to weight the RCT sample to achieve balance in the moderators observed in both samples: race and baseline number of cigarettes per day.
Exploring two methods of estimating the conditional probability of being in the RCT sample (logistic regression and generalized boosted models) and two sets of variables (moderators only and moderators plus other covariates), we find that the weights generated from the logistic sample membership based on the combination of moderators and covariates observed in both samples (race, baseline cigarettes per day, age and gender) result in the best balance, for the moderators as well as the other covariates.
Specifically, these weights reduced the standardized mean differences (between the target population sample and the RCT sample) for all race categories, age and gender to under 0.05; and reduced that for baseline daily number of cigarettes from 0.43 before weighting to 0.15 after weighting.

Using these weights, we fit the same outcome model with interactions to the weighted RCT sample; results are presented in Table \ref{table:3}. The coefficients of the interaction terms are similar to those from the unweighted model in Table \ref{table:2}, but their confidence intervals are wider due to the weighting. Applying the weighted-outcome-model-based sensitivity analysis, we have TATE ranging from $-8.36$, 95\% CI $(-12.50,-4.23)$ (corresponding to mean baseline addiction score 3) to $-4.83$, 95\% CI $(-8.76,-0.90)$ (corresponding to mean baseline addiction score 5). %results in the range and 95\% confidence bound presented in the bottom plot in Figure \ref{figure:dataexample}.

Figure \ref{figure:dataexample} includes the ranges of TATE with confidence bounds from both the sensitivity analyses presented here. These ranges are below zero, suggesting that this smoking cessation intervention, if applied to the target population, would result in smoking reduction. These results are generally consistent with the RCT findings, but give us more confidence in what the effects would be among the target population of cigarette smokers among people seeking alcohol/drug use treatment in the US.

\section{Discussion}\label{sxn:discuss}

This paper presents approaches for assessing the sensitivity of the estimated treatment effect to an unobserved treatment effect moderator when generalizing from a RCT to a target population. 
The paper addresses two distinct situations.
The first situation arises naturally as researchers find from analyzing RCT data that there are variables that moderate treatment effect, but some of those variables ($V$) are not available from the data they have for the target population.
For this case, we offer an outcome-model-based method, a method fully based on weighting, and a weighted-outcome-model-based method which combines elements of the first two. The two methods based on the outcome model are relatively straightforward and only requires specifying a plausible range for the mean of $V$ in the target population; the weighting-based method is complicated and thus is recommended only for the case with binary moderators.

%The purpose of sensitivity analysis in this $V$ case is to assess the plausible range for  TATE given some plausible range we specify about the unobserved moderator in the target population.
The second situation is when researchers are concerned that there may be treatment effect heterogeneity that is not detected given the observed variables in the RCT, and ask what that implies about TATE.
Here we consider a composite variable $U$ that represents the remaining effect moderation factors after accounting for the observed moderators (and possibly other covariates).
For this case, we offer a bias-formula-based method and a weighting-plus-bias-formula-based method.
With both methods, we vary two parameters, one representing $U$'s association with being in the RCT, the other representing its moderation effect, and determine how TATE estimates change as a function of these parameters.

In this paper, we consider the RCT and the target population samples as disjoint sets.
The proposed methods, however, are easily adapted to the situation where the RCT sample is a subset of the target population sample.
In that case $S=1$ denotes being in the RCT, but all individuals (with $S=1$ or $S=0$) are in the target population sample.
All quantities regarding the target population are not conditioned on $S=0$.
The weighting procedures involve modeling $S$ using the target population sample (which includes RCT subjects), and weighting RCT subjects by the inverse of their probability of being in the RCT.

There are several directions for future extension of these sensitivity analyses. 
First, these methods apply when we do not have outcome data for the target population. %, and need to rely on the assumption of sample ignorability of treatment effect conditional on the moderators (both those that are observed and the unobserved one being considered).
There are, however, situations where the outcome under control or a combination of outcome under control and treatment for different individuals is observed in the target population.
Currently, methods exist that use target population outcome data under control, but only for generalization where all data, including moderating variables, are observed \citep{Kern2016a}.
The proposed methods could be adapted to incorporate target population outcome data.

Second, the proposed methods that use weighting are based on a specific method of adjusting treatment effect estimates for the differential distribution of a moderator---adjustment by weighting.
Another adjustment strategy is to fit a flexible model of the outcome as a function of treatment, covariates and interaction terms, and either impute potential outcomes for individuals in the target population as the basis to estimate  TATE, or estimate treatment effects for covariate strata and average these estimates using the target population covariate distribution \citep*[see][for example]{Kern2016a}.
Future work should investigate ways to extend the sensitivity analysis procedures we present here to that approach.

Third, the proposed methods do not cover the case where researchers are concerned about a specific variable (e.g., parents' education attainment) that is known or suspected (based on prior evidence) to moderate treatment effect, but is not measured by the RCT. Methods for $U$ do not apply as the moderator in this case is a specific variable, not the remaining composite moderator. This situation is related to our first situation with a specific moderator $V$, except that this variable is not measured in the RCT. Future investigation is needed to extend our first set of proposed methods to this situation, perhaps using a combination of additional assumptions and external information about this variable.

Fourth, the proposed outcome-model- and bias-formula-based methods assume a linear model for the potential outcomes. In the case of a binary outcome, for example, this means assuming treatment affects the outcome on the risk difference scale. Recent work by \citet{Ding2015a} shows that for a binary outcome, effects measured on the odds ratio scale tend to be less heterogenous than on the risk difference (and also risk ratio) scale. One of our next steps is to adapt these sensitivity analysis methods to effect scales that are less heterogeneous for specific outcomes, for example odds ratio for a binary outcome and rate ratio for a count outcome.

To conclude, in this paper we have presented methods for assessing sensitivity of results regarding generalizability of treatment effects to effect heterogeneity on unobserved characteristics.
These methods are helpful to researchers and policy makers who are interested in the effect of a treatment for a certain population, given the common situation that not all effect moderators are measured both in the RCT and in the target population.

\bibliographystyle{plainnat}
\bibliography{Generalization}

\appendix

\section*{Acknowledgements}
TQN is supported by NIDA grant T32DA007292 (PI: Renee M. Johnson). EAS and CE are supported by NSF grant DRL-1335843. We thank Ryoko Susukida and Ramin Mojtabai for their kind support for our search for an appropriate data example.%; and thank the Area Editor, the Associate Editor and the Reviewer for their very helpful comments on the manuscript.

\section*{Supplement A: R-code for the simulation study}

\section*{Supplement B: R-code for implementing the outcome-model-based and weighted-outcome-model-based methods on the data example}

\end{document}